\input ppltex.sty
\input abbrev.sty
\def\z{\zeta}
\let\xnewsec=\newsec
\def\newsec#1{\xnewsec F{#1}}

\hsize 4.85in \hoffset 0.825in
\def\figno#1{}

\null
\vskip 0.6in
\ctrline{\titl Velocity-Space Diffusion in a Perpendicularly}
\ctrline{\titl Propagating Electrostatic Wave}
\vskip 20 pt
\ctrline{Charles F. F. Karney}
\vskip 5 pt
\ctrline{Plasma Physics Laboratory, Princeton University}
\ctrline{Princeton, New Jersey 08544, U.S.A.}
\vskip 40 pt
\newsec{Abstract}

The motion of ions in the fields $\vec{B} = B_0 \^{\vec{z}}$ and
$\vec{E} = E_0 \^{\vec{y}} \cos(k_\perp y - \omega t)$
is considered.  When $\omega \grgr \Omega_i$ and $v_\perp \grapprox
\omega/k_\perp$, the equations of motion may be reduced to a set
of difference equations.  These equations exhibit stochastic behavior
when $E_0$ exceeds a threshold.  The diffusion coefficient above the
threshold is determined.  Far above the threshold, ion Landau damping is
recovered.  Extension of the  method to include parallel propagation
is outlined.

\vfill\noindent
Presented at the International Workshop on Intrinsic Sto\-chas\-ti\-city
in Plasmas, Carg\`ese, Corsica, France, June 18--23, 1979.  Published in
{\it Intrinsic Stochasticity in Plasmas}, edited by G. Laval and
D. Gr\'esil\-lon (Editions de Physique Courtaboeuf, Orsay, 1979), pp.\
159--168.

\par\eject\newsec{Equations of Motion}
% \newseca{\vjust{\vjust to 1.7in{\vfill}\hjust{Equations of Motion}}}

Consider an ion in a uniform magnetic field and a perpendicularly
propagating electrostatic wave,
$$\vec{B} = B_0\^{\vec{z}},\qquad
\vec{E} = E_0\^{\vec{y}} \cos(k_\perp y - \omega t). \eqn{\en{fields}{1}}$$
\xdef\fields{\eqprefix {1}}%
Normalizing lengths to $k_\perp^{-1}$ and times to $\Omega_i^{-1}$
($\Omega_i = q_i B_0/m_i$), the Lorentz force law for the ion becomes
$$\" y + y = \a \cos(y - \nu t),\qquad
\.x = y, \eqn{\en{lfl}{2}}$$
\xdef\lfl{\eqprefix {2}}%
where
$$\eqalignno{\nu &= \omega/\Omega_i, &(\en{pdef}{3}a)\cr\eqskp
	\a &= {E_0/B_0\over \Omega_i/k_\perp}. &(\+b)\cr}$$

We solve (\lfl) by approximating the force
due to the wave by impulses at those points where the
phase is slowly varying, i.e., at $\.y = \nu$.  The
trajectory of the ion is given in Fig.\ 1.  Expanding
the trajectory about the resonance point, we find that the magnitude of the
impulses is given by
$$\eqalignno{B &= \int_{-\inf}^\inf \a \cos(\phi_c - \half y_c t^{\prime 2})
dt^\prime \cr\eqskp
&= \a (2\pi/\abs{y_c})^{1/2} \cos[\phi_c - \sign(y_c)\pi/4],
&(\en{Beq}{4})\cr}$$
where $\phi_c = y_c - \nu t_c$ and $t_c$ and $y_c$ are
the time and position of the wave-particle ``collision.''  We may determine
the Larmor radius and phase of the ion at the end of the $j$th orbit
[the beginning of the $(j + 1)$th orbit] in terms of these
quantities at the beginning of the $j$th orbit.  (Details are
given in Ref.\ 1.)  The resulting difference equations are
$$\vjust{\eqaligntwo{u &= \t - \r ,&  v &= \t + \r ,& (\en{de}{5}a)\cr\eqskp
\t &= \half(v + u) ,&  \r &= \half(v - u) ,& (\+b)\cr}
\eqskip
\eqalignno{u_{j + 1} - u_j &= 2\pi\d - 2\pi A \cos v_j ,& (\+c)\cr\eqskp
v_{j + 1} - v_j &= 2\pi\d + 2\pi A \cos u_{j+1} ,& (\+d)\cr}}$$
\xdef\de{\eqprefix {5}}%
where
$$\eqalignno{\t &= \nu t \mod{2\pi},&(\en{vdef}{6}a)\cr\eqskp
\r &= (r^2 - \nu^2)^{1/2} - \nu \cos^{-1}(\nu/r) + \nu\pi -
\pi/4,& (\+b)\cr\eqskp
A &= \left( {2\over \pi} \right) ^{1/2}
{\a\nu(r^2 - \nu^2)^{1/4}\over r^2} ,&(\en{ddef}{7}a)\cr\eqskp
\d &= \nu - n .& (\+b)\cr}$$
\xdef\ddef{\eqprefix {7}}%
Here $r$ is the normalized Larmor radius, $k_\perp v_\perp/\Omega_i$,
and $n$ is an integer.  The limits of validity of (\de) are
$$\nu \grgr 1,\qquad r-\nu \grgr \nuth,\qquad
A \lsls (r^2-\nu^2)^{3/2}/r^2 . \eqn{\en{foo}{8}}$$

In Fig.\ 2 we compare the trajectories obtained using the exact
equations of motion, (\lfl), with those obtained from the
difference equations, (\de).  We see that the agreement is
very good indicating that (\de) is an excellent approximation
of (\lfl).

There are three advantages to using the difference equations in pref\-er\-ence
to the Lorentz force law.  Firstly, they are much quicker to 
solve numerically.  Secondly, because of the way the equations were derived,
the results are easier to interpret.  Lastly, the equations have
separated out two velocity-space scales, the $\r$ scale
($\sim \Omega_i/k_\perp$) and the $r$ scale ($\sim \omega/k_\perp$).  We therefore
treat $A$ which is a function of $r$ as a constant
when iterating the equations.  This means that the diffusion coefficient
is independent of $\r$ and so is much easier to determine numerically.

\newsec{Examples of Trajectories}

When $A$ is infinitesimal, (\de) may be solved by integrating
(summing?)\ over unperturbed orbits.  Substituting
$u_j = u_0 + 2\pi\d j$ and $v_j = v_0 + 2\pi\d j$ into the
right hand sides of (\de c) and (\de d) gives
$$\eqalignno{\r_N - \r_0 &= 2\pi A \cos(\r_0 - \pi\d)\cr\eqskp
&\qquad\null\times\case{(-1)^{\d}\cos\t_0\,N, &
\for \d = \hjust{integer},\cr\caseskp
{\sin(2\pi\d N + \t_0) - \sin\t_0 \over 2\sin(\pi\d)}, &
\for \d \ne \hjust{integer}.\cr} &(\en{rsec}{9})\cr}$$
\xdef\rsec{\eqprefix {9}}%
Note that the trajectory is secular or not depending on
whether or not $\d$ is an integer ($\omega$
is a cyclotron harmonic).  Formally, we may compute a diffusion
coefficient using
$$\D =  \Lim_{N \to \inf}{\ave{(\r_N - \r_0)^2}\over 2N} .
\eqn{\en{ddefm}{10}}$$
\xdef\ddefm{\eqprefix {10}}%
Substituting (\rsec) into (\ddefm) gives
$$\D = \pi^2 A^2 \cos^2\r_0 \sum_{m = - \inf}^\inf \^\d(\d - m),
\eqn{\en{qldn}{11}}$$
where $\^\d$ is the Dirac delta function.  Converting back to $r$
and $t$ and undoing the normalizations gives
$$D = {\pi\over 2}{q_i^2E_0^2\over m_i^2} \bigglp {\omega\over k_\perp v_\perp}
\biggrp ^2 \sum_{m=-\inf}^\inf J_m^2\bigglp {k_\perp v_\perp\over
\Omega_i}\biggrp \,\^\d(\omega - m\Omega_i), \eqn{\en{qld}{12}}$$
which is the usual quasi-linear diffusion coefficient.

If we consider finite but small $A$, then all the
trajectories are bounded.  There are three distinct cases,
$\d = 0$ (which is the case considered by Fukuyama 
{\ital et al.}\ref{2}), $\d = \half$, and
$\d \ne 0$ or $\half$.  The trajectories for these cases are
shown in Fig.\ 3.

When $A$ is increased, the system undergoes a stochastic
transition, an example of which is shown in Fig.\ 4 for
$\d = 0.23$.  Below the stochasticity threshold, nearly
all the trajectories are integrable [Fig.\ 4(a)] or, if there
are stochastic trajectories, they are bounded in $\r$ [Fig.\ 4(b)].  Above
the threshold, nearly all the trajectories are stochastic and unbounded.  The
value of the threshold may be numerically determined and is found to be
$A = A_s = \quarter$.  Above this value of $A$, the kick
received by the ion during one transit through resonance
is sufficient to change the phase of the kick
received when next in resonance by $\pi/2$.

\newsec{Diffusion Coefficient}

When computing the diffusion coefficient numerically, it is convenient
to work with the correlation function, $C_k$, where
$$C_k = \ave{a_ja_{j+k}}, \eqn{\e}$$
$a_j$ is the particle acceleration, $a_j = \r_{j+1} - \r_j$,
and the average is over an ensemble of particles and over the length
of a given trajectory (i.e., over $j$).  Then the diffusion coefficient,
$\D$, is given by
$$\D = {1\over 2} C_0 + \sum_{k=1}^{\inf} C_k . \eqn{\en{Ddef}{14}}$$
\xdef\Ddef{\eqprefix {14}}%
[This definition is equivalent to (\ddefm).]  The advantages of defining
$\D$ in this way are twofold.  Firstly, the statistical fluctuations
in the computation are minimized.  Secondly, it is easy to introduce
the effects of collisions on the diffusion coefficient.  This
is accomplished as follows:  If $k_0$ is the
mean number of cyclotron periods between decorrelating collisions
(such collisions need only result in deflection by $1/\nu$,
a small angle), then the probability of such a collision taking place
in $k$ periods is $1-\exp(-k/k_0)$ since collisions are
independent events.  Collisions may
then be included in the computations of $\D$
by replacing $C_k$ in (\Ddef) by $C_k \exp(-k/k_0)$.

In the limit $A \grgr A_s$, the kicks the ion receives are
uncorrelated so that only $C_0$ is nonzero.  Assuming that
the trajectory is ergodic, we obtain $\D = \half \pi^2 A^2$.  When
$A$ is not large, we account for the correlations between the kicks
received by the ion by  writing
$$\D = \half \pi^2 A^2 g^2(A), \eqn{\en{Gdef}{15}}$$
\xdef\Gdef{\eqprefix {15}}%
Numerically determining $g(A)$ we find that approximately
$$g(A) = \max(1-A_s^2/ A^2, 0),\eqn{\en{geq}{16}}$$
with $A_s = \quarter$.

Converting (\Gdef) back into usual variables we obtain
$$D = {1\over 2}{q_i^2E_0^2\over m_i^2} \bigglp {\omega\over k_\perp v_\perp}
\biggrp ^2 {g^2(A)\over(k_\perp^2v_\perp^2-\omega^2)^{1/2}}.
\eqn{\en{sld}{17}}$$
In the limit $A \grgr A_s$, when $g(A) \approx 1$, this is just the
zero-magnetic-field result,
$(\pi/2) (q_i/m_i)^2 E_0^2 \,\^\d(\omega - k_\perp v_y)$, averaged
over Larmor phase.  Thus, in this limit, we
recover ion Landau damping.

\newsec{Extension to Parallel Propagation}

The diffusion coefficient was so easily calculated above
because of the simp\-li\-fi\-ca\-tion obtained by
reducing the problem to difference
equations, (\de).  This reduction may be achieved in similar
problems.  We consider here the case where the wave has some component
of parallel propagation so that (\fields) becomes
$$\vec{B} = B_0\^{\vec{z}},\qquad
\vec{E} = E_0(\^{\vec{y}} + k_\para\^{\vec{z}}/k_\perp)
\cos(k_\perp y + k_\para z - \omega t).
\eqn{\en{pfields}{18}}$$
Adopting the same normalization as before, we obtain
$$\eqalignno{\" y + y &= \a \cos(y + \z z - \nu t),&(\en{plfl}{19}a)\cr\eqskp
\" z &= \a\z\cos(y +\z z -\nu t),& (\+b)\cr}$$
\xdef\plfl{\eqprefix {19}}%
where $\z = k_\para/k_\perp$.  The difference equations for a particle
with normalized Larmor radius, $r$, and normalized parallel velocity,
$w = k_\perp v_\para/\Omega_i$, are
$$\vjust{\eqaligntwo{u &= \t - \r ,&  v &= \t + \r ,& (\en{pde}{20}a)\cr}
\eqskip
\eqalignno{u_{j + 1} - u_j &= 2\pi\gamma_{j+1/2} - 2\pi A \cos v_j ,&
(\+b)\cr\eqskp
v_{j + 1} - v_j &= 2\pi\gamma_{j+1/2} + 2\pi A \cos u_{j+1} ,&
(\+c)\cr\eqskp
\gamma_{j+1/2} &= \d - \beta(\r_j +\pi A\cos v_j) ,&(\+d)\cr}}$$
\xdef\pde{\eqprefix {20}}%
where the variables $\t$ and $\r$ are given by
$$\eqalignno{\t &= \nu t - \z z \mod{2\pi},&(\en{pvdef}{21}a)\cr\eqskp
\r &= (r^2 - \mu^2)^{1/2} - \mu \cos^{-1}(\mu/r) + \mu\pi -
\pi/4 - 2\pi m. &(\+b)\cr}$$
The definitions of the parameters $A$, $\beta$, and $\r$ are
$$\eqalignno{A &= \left(2\over\pi\right)^{1/2}
{\a\mu Q\over r(r^2-\mu^2)^{1/4}},
&(\en{ppdef}{22}a)\cr\eqskp
\beta &= \z^2r/(\mu Q),&(\+b)\cr\eqskp
\d &= \mu + \beta\r-n.&(\+c)\cr}$$
\xdef\ppdef{\eqprefix {22}}%
Here, $m$ and $n$ are integers and
$$\eqalignno{\mu &= \nu -\z w,&(\e a)\cr\eqskp
Q &= {(r^2-\mu^2)^{1/2}\over r} - \left[\pi-\cos^{-1}
\left({\mu\over r}\right)\right]{\z^2 r\over \mu}.&(\+b)\cr}$$
Despite appearances $\d$ is a parameter independent of $\r$
since the quantity $\mu + \beta \r$ is a constant.  (This
follows from energy conservation in the wave frame.)  The restrictions
on the validity of (\pde) are
$$\mu \grgr 1,\qquad r-\mu \grgr  (\half\mu)^{1/3}. \eqn{\e}$$

A comparison between the exact equations of motion, (\plfl),
and the difference equations, (\pde), is shown in Fig.\ 5 for $\z = 1$
(propagation at $45\deg$).  Again, there is excellent agreement.

The results of Smith and Kaufman\ref{3} may be obtained in the limit
$\beta \to \inf$ and $A \to 0$ ($Q \to 0$).  In that
case, the change in $\r$ is negligible so that it is necessary to rescale
the velocity variable by defining
$$\sigma_j = 4\pi\beta\r_j - 2\pi\d.\eqn{\e}$$
Equation (\pde) then becomes
$$\eqalignno{
\t_{j+1}-\t_j &= -\sigma_j - \half K\cos(\t_j+\r),&(\en{smm}{26}a)\cr\eqskp
\sigma_{j+1}-\sigma_j &= \half K[\cos(\t_{j+1} -\r)
+\cos(\t_j+\r)],&(\+b)\cr}$$
where $K = 4\pi^2 A\beta$.  In (\+) $\r$ is a constant.  Setting
$$\psi = \sigma + \half K cos(\t + \r)\eqn{\e}$$
gives
$$\eqalignno{
\t_{j+1} - \t_j &= - \psi_j,&(\en{sm}{28}a)\cr\eqskp
\psi_{j+1} - \psi_j &= K\cos\r\cos\t_{j+1}.&(\+b)\cr}$$
This is the ``standard mapping'' studied by Chirikov.\ref{4}  The
island overlap condition for this mapping is
$\abs{K\cos\r} > \pi^2/4$ or
$$\abs{\a} > \abs{16\z^2J_\mu(r)}^{-1},\eqn{\e}$$
which is the stochasticity threshold obtained by Smith and Kaufman.  The
island overlap criterion is a significant overestimate of the
stochasticity threshold for the standard mapping.\ref{4}  Greene\ref{5}
has calculated that the true threshold is a factor $(\pi^2/4)/0.971635
\approx 2\half$ smaller than the result given above.

\newsec{Acknowledgments}

The author wishes to thank N. J. Fisch, J. M. Greene,
J. A. Krommes, and A. B. Rechester for useful discussions.

This work was supported by the U. S. Department of Energy
under Contract No.\ EY--76--C--02--3073.

\newsec{References}

\refno{1}C. F. F. Karney, Phys.\ Fluids \vol{21}, 1584 (1978);
Phys. Fluids \vol{22}, 2188 (1979).
\refno{2}A. Fukuyama, H. Momota, R. Itatani, and T. Takizuka,
Phys.\ Rev.\ Lett.\ \vol{38}, 701 (1977).
\refno{3}G. R. Smith and A. N. Kaufman,
Phys.\ Rev.\ Lett.\ \vol{34}, 1613 (1975); Phys.\ Fluids \vol{21}, 2230 (1978).
\refno{4}B. V. Chirikov, Phys.\ Repts.\ \vol{52}, 265 (1979).
\refno{5}J. M. Greene, J. Math.\ Phys.\ \vol{20}, 1183 (1979).

\vfill\penalty-10000
%\newseca{Figures}

\Fig. 1.  (792078:0.6)
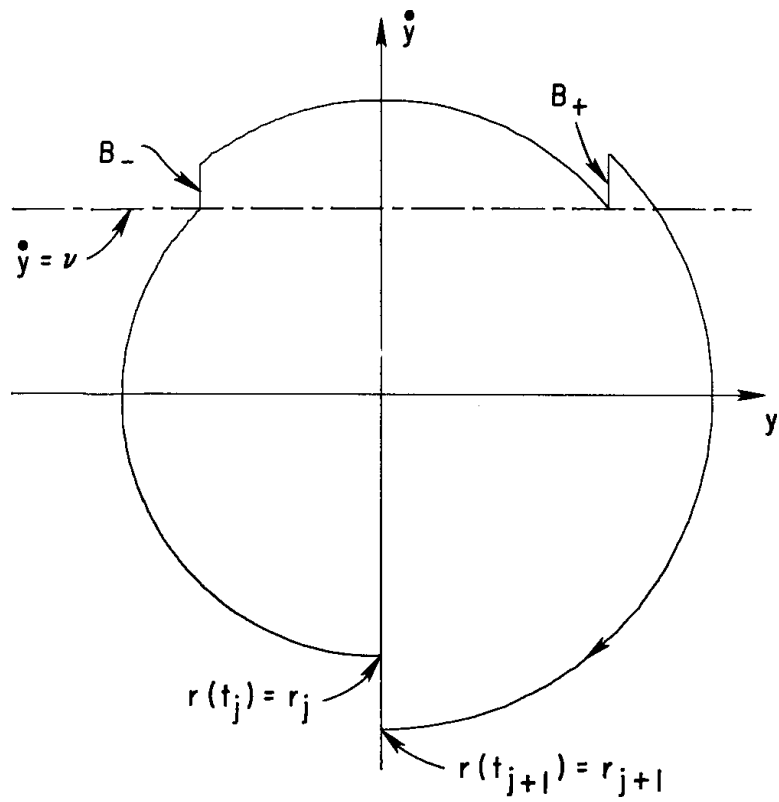
Motion of an ion in velocity space, showing the kicks it
receives when passing through wave-particle resonance.  \figno{792078}
\Fig. 2.  (792087:0.8)
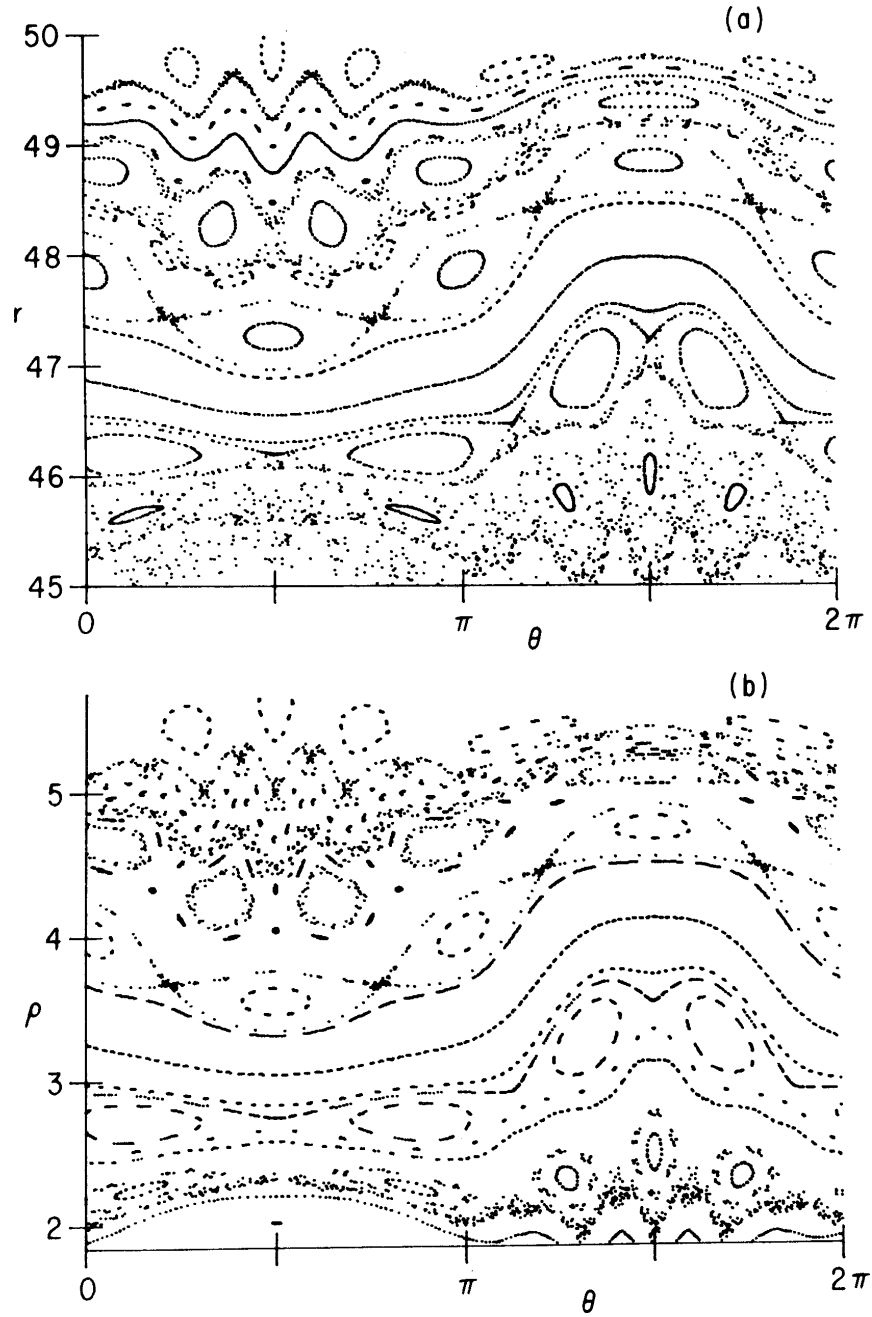
Comparison of the difference equations with the Lorentz
force law.  (a) Trajectories computed using (\lfl)
with $\nu = 30.23$ and $\a = 2.2$.  (b) Trajectories computed
using (\de) with $\d = 0.23$ and $A = 0.1424$, which are given by (\ddef)
with $\nu = 30.23$, $\a = 2.2$, and $r  = 47.5$.  In each case the trajectories
of 24 particles are followed for 300 orbits.  \figno{792087}
\Fig. 3.  (792080:0.8)
Trajectories for small $A$ and (a) $\d=0$, (b) $\d=\half$,
and (c) $\d\ne 0$ or $\half$.  \figno{792080}
\Fig. 4.  (792233:1.25)
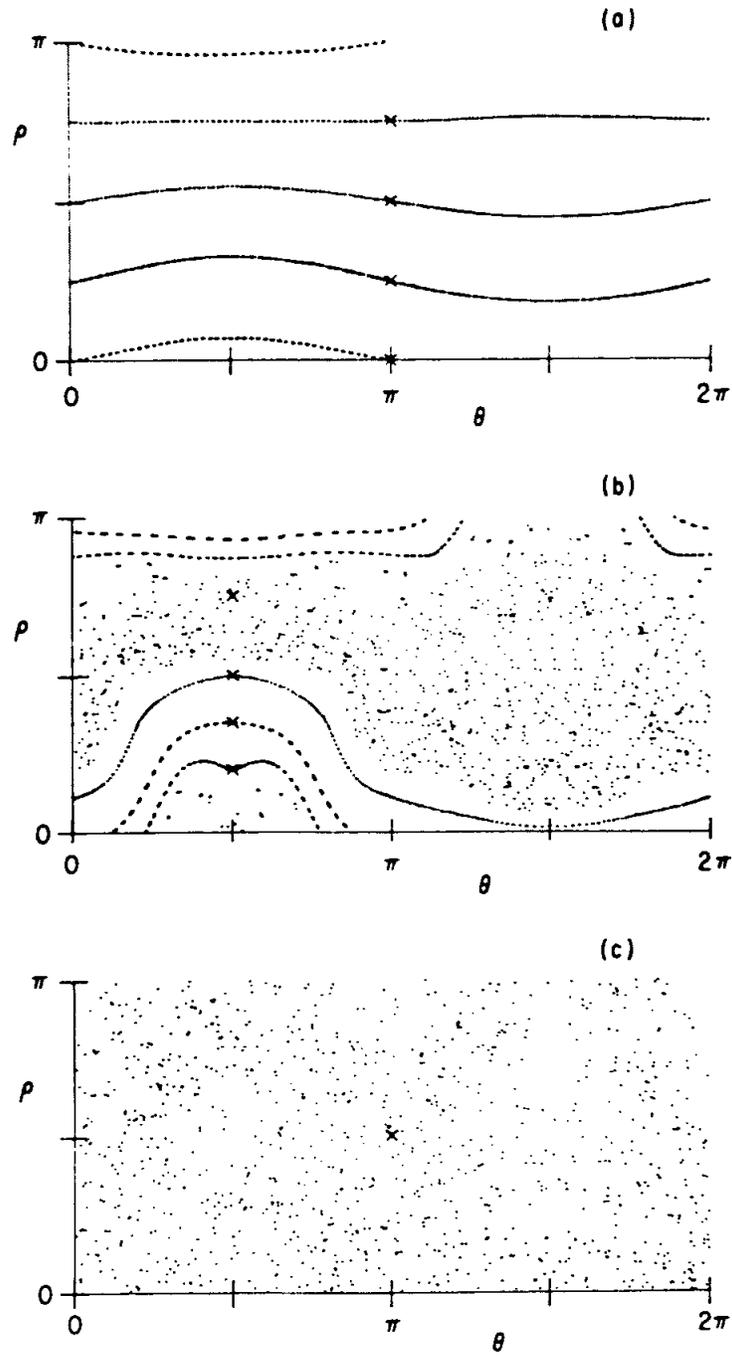
The trajectories of ions with
$\d = 0.23$ and (a) $A=0.05$, (b) $A=0.2$, (c) $A=0.35$.  The initial positions
are shown by crosses.  \figno{792233}
\Fig. 5.  (792234:1.33333)
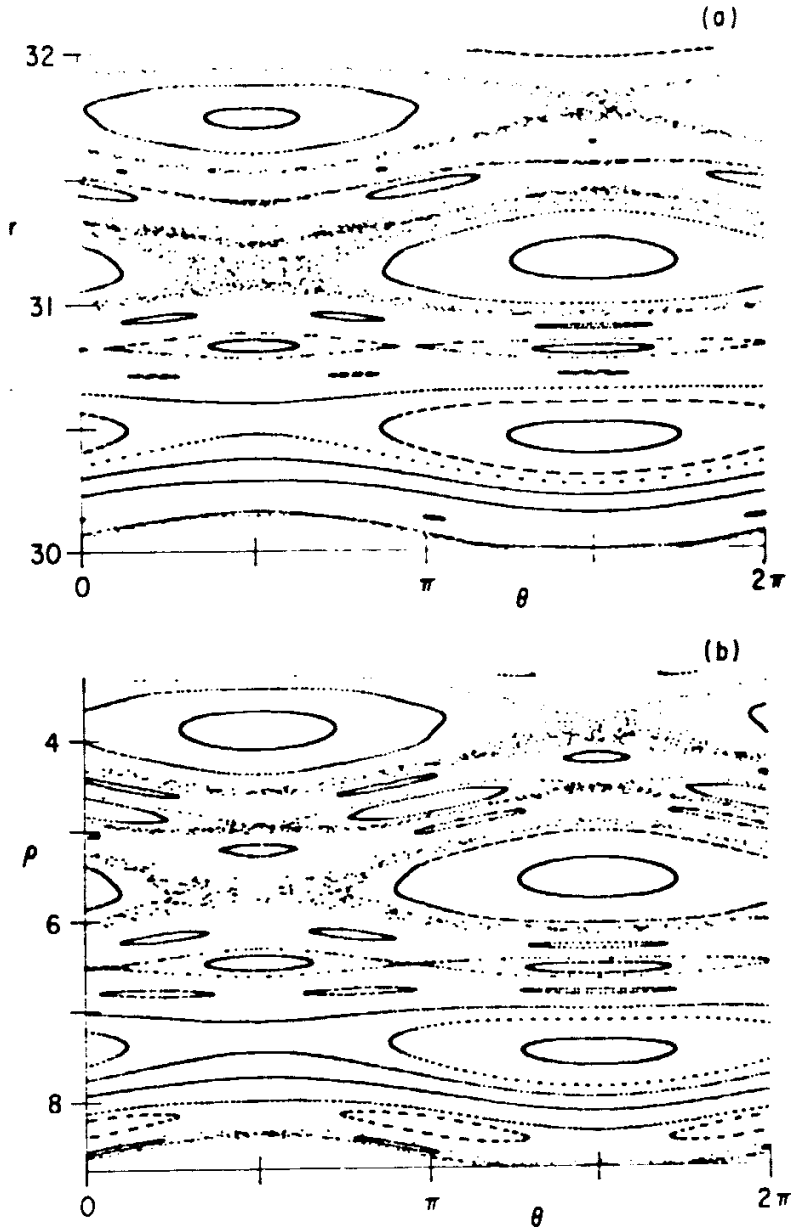
Comparison of the difference equations with the Lorentz
force law for finite $k_\para$.  (a) Trajectories computed using (\plfl)
with $\nu=20.23$, $\a = 0.23$, and $\z = 1$.  The total energy
($\hjust{kinetic} + \hjust{electrostatic}$) of each particle
is the same and is chosen so that
$w\approx 0$ when $r = 31$.   (b) Trajectories computed
using (\pde) with $\d = 0.8548$, $A=-0.06768$, and $\beta=-0.5595$
which are given by (\ppdef) with $\nu = 20.23$, $\a = 0.23$,
$\z=1$, $r = 31$, $w=0$, and $m=10$.  In each case the
trajectories of 24 particles are followed for 300 orbits.  The vertical
scale is inverted in (b) since $Q < 0$.  \figno{792234}
\bye